\input{aipcheck}

\documentclass[
    ,final            
  ]
  {aipproc}

\layoutstyle{6x9}
\usepackage{bm,amsmath}

\newcommand\npb[3]{{\it Nucl. Phys. }{\bf B #1} (#2) #3}

\newcommand\prd[3]{{\it Phys. Rev. }{\bf D #1} (#2) #3}

\begin{document}

\title{Nonlinear evolution equations in QCD and effective Hamiltonian  at high energy}


\author{Anna M. Sta\'sto}{
  address={Physics Department, Brookhaven National Laboratory, Upton, NY 11973, USA\\
  and\\
  H.~Niewodnicza\'nski Institute of Nuclear Physics,  31-342 Krak\'ow, Poland}
}

\begin{abstract}
In this talk I briefly present  recent developments in the theory of  the Color Glass Condensate.
  The duality between the dense and dilute regimes of the gluon field is discussed as well as the effective selfdual Hamiltonian which includes both Pomeron merging and Pomeron splitting.
\end{abstract}

\maketitle


\subsection{Introduction: high energy limit and the Pomeron}

One of the central problems in the theory of strong interactions  is the understanding 
of the behaviour of the hadronic cross sections in the limit of high energies.
The experimental data on the total cross section show slow but distinct increase with energy.
It has been suggested that this rise, which can be parametrized by a power-like form,
\begin{equation}
\sigma(s) \; \sim \;  s^{0.08} \; ,
\label{eq:softp}
\end{equation}
 is mediated by the 'soft Pomeron' \cite{DL}.  Perturbative calculation summing the leading logarithmic contributions of large logs $\ln s/t$ results in the BFKL Pomeron, which however leads to a stronger rise
 of the cross section with energy \cite{BFKL}
 \begin{equation}
 \sigma(s) \sim s^{4\ln2\, \alpha_s N_c/\pi} \; ,
 \end{equation}
with the leading exponent being much larger than in (\ref{eq:softp}) when evaluated at physically interesting values of the strong coupling constant ($\alpha_s \simeq 0.2$).
In any of these cases, the power-like  increase of the cross section is in contradiction with
the Froissart bound \cite{Froissart} which allows at most logarithmic increase 
with energy
\begin{equation}
\sigma(s) \sim \sigma_0 \ln^2 s \; ,
\end{equation}
with a  normalization coefficient $\sigma_0$ related to the inverse of the pion mass squared.  Froissart bound stems from the  very general principles: unitarity and the finite range of the strong interactions. Thus, the important question arises whether it is possible to identify and calculate other type of Feynman diagrams which will lead  to the restoration of the unitarity. 

\subsection{Color Glass Condensate and the JIMWLK equation}
The Color Glass Condensate  is an effective theory of the strong interactions at very high energies.
Apart from the standard BFKL evolution of the gluon density  it also  contains the recombination diagrams which are important when the gluon density becomes very high.
It therefore describes  the phenomenon known as perturbative parton saturation. When the density of gluons
becomes very high due to their enhanced splitting described by the BFKL Pomeron, the recombination effects start to become important and reduce the growth.
It is believed that this phenomenon is important for the restoration of the unitarity.
The Color Glass Condensate theory describes the scattering of the projectile off a target which constitutes a dense system of soft gluons.
The S-matrix for the scattering of the quark-antiquark dipole in the target field  $\alpha$ is described by
\begin{equation}
S(\bm{x},\bm{y}) \; = \; \frac{1}{N_c} {\rm Tr} (V_{\bm{x}}V^{\dagger}_{\bm{y}} ) \; ,
\label{eq:smatrix}
\end{equation}
where Wilson line
\begin{equation}
V_{\bm{x}} \; = \; P \exp \big(ig \int dx^- \alpha_a(x^-,\bm{x}) \, t^a\big) \; ,
\end{equation}
is the path ordered exponential along the trajectory of the projectile. The physical scattering matrix between the $q\bar{q}$
dipole and the hadron target is  obtained by taking the average of (\ref{eq:smatrix}) over the color field of the target
\begin{equation}
\label{eq:savg}
\langle S(\bm{x},\bm{y})\rangle_{\tau} \; = \; \int D[\rho] \, {\cal  Z}[\rho]_{\tau} \, S(\bm{x},\bm{y}) \; ,
\end{equation}
where $\rho$ is a color charge which generates field $\alpha$. One cannot compute the weight function ${\cal Z}[\rho]$
since it contains nonperturbative information about the hadron but one can compute the evolution of 
this weight function  with increasing rapidity $\tau \sim \ln s$.
The basic equation of the Color Glass Condensate theory is the 
JIMWLK equation \cite{JIMWLK} which governs the evolution of the weight function ${\cal Z}[\rho]_{\tau}$
with rapidity
\begin{equation}
\label{eq:JIMWLK}
\frac{\partial {\cal Z}[\rho]_{\tau}}{\partial \tau} \; = \; H_{JIMWLK} \, {\cal Z}[\rho]_{\tau}  \; ,
\end{equation}
where $H$ is the JIMWLK Hamiltonian. The JIWMLK equation is a very complicated functional evolution equation, which however reduces  to one closed and relatively simple equation, the Balitsky-Kovchegov equation, in the large $N_c$ limit and in the dipole picture.   
Diagramatically JIMWLK equation contains BFKL Pomeron ladder diagrams as well as the 
triple Pomeron interaction diagrams which reduce the growth of the gluon density.
 More precisely it contains Pomeron merging diagrams, however it misses the Pomeron splittings.
 One can understand it by looking at the structure of the operator  $H_{ JIMWLK}$
 \begin{equation}
 \label{eq:hjimwlk}
H_{ JIMWLK} \; = \; \frac{1}{2\pi} \, \int_{\bm{x},\bm{y},\bm{z}} \, K_{\bm{xyz}}  \, \frac{\delta}{\delta \alpha^a(\bm{x})} \, 
\left[   1+\tilde{V}_{\bm{x}}^{\dagger}\tilde{V}_{\bm{y}}-\tilde{V}_{\bm{z}}^{\dagger}\tilde{V}_{\bm{y}}-\tilde{V}_{\bm{x}}^{\dagger}\tilde{V}_{\bm{z}}\right]^{ab}
\,  \frac{\delta}{\delta \alpha^b(\bm{y})} \; ,
 \end{equation}
 which is second order in derivatives $\frac{\delta}{\delta \alpha}$ and all orders (through $\tilde{V}$'s)\footnote{$\tilde{V}$ denotes the Wilson line in the adjoint representation.} in the field $\alpha$. That means, that the evolution of the correlation functions in the field $\alpha$ can be 
 (by using (\ref{eq:savg},\ref{eq:JIMWLK},\ref{eq:hjimwlk})) schematically represented as
 \begin{equation}
 \frac{\partial \stackrel{n}{\overbrace{\langle\alpha\dots\alpha\rangle}}}{\partial \tau} \, = \, \sum_{m \ge n} {\cal K}_m \otimes
 \stackrel{m}{\overbrace{\langle\alpha\dots\alpha\rangle}} \; .
 \end{equation}
  The evolution of $n$ point functions is coupled to the  higher order correlation functions
  but not to the lower correlation functions. 
  \subsection{Dual version of JIMWLK and effective Hamiltonian}
  In order to include the additional diagrams corresponding to the Pomeron splittings in the Color Glass Condensate formalism,  terms which are higher order in derivatives have to be incorporated  \cite{MSW}.  By including these terms in the evolution, the   Pomeron loops are generated, which are known 
  to be important contributions at very high energy. It has been suggested \cite{KL}, that in general there is a duality relation between the evolution of the dense system, governed by the JIWMLK equation, and the dilute regime. In particular, the Pomeron mergings and splittings are believed to be the dominant
  diagrams in the dense and the dilute regimes, respectively, of the gluon field.
  In particular the evolution in the dilute regime is governed by the dual equation to the JIMWLK evolution.
  Formally, the duality can be expressed as the transformation \cite{KL,HIMST}
  \begin{equation}
  x^- \leftrightarrow x^+, \quad \frac{\delta}{\delta \alpha^a(x^-,\bm{x})} \leftrightarrow i\rho^a(x^+,\bm{x}), \quad
  \alpha^a(x^-,\bm{x}) \leftrightarrow -i \frac{\delta}{\delta \rho^a(x^+,\bm{x})} \; ,
  \end{equation}
  where $\rho$ is the charge of the target and the $\alpha$ is the field generated by this charge measured by the projectile.
  The dual version of the JIMWLK equation is then
  \begin{equation}
  \label{eq:brem}
  \frac{\partial {\cal Z}[\rho]_{\tau}}{\partial \tau} \; = \; \tilde{H}\, {\cal Z}[\rho]_{\tau}  \; ,
  \end{equation}
  where the dual Hamiltonian
  \begin{equation}
  \label{eq:hbrem}
  \tilde{H} \; = \;  -\frac{1}{2\pi} \int_{\bm{x},\bm{y},\bm{z}} \, K_{\bm{xyz}} \;
  {\rho^a(\bm{x})} \, \left[   1+\tilde{W}_{\bm{x}}^{\dagger}\tilde{W}_{\bm{y}}-\tilde{W}_{\bm{z}}^{\dagger}\tilde{W}_{\bm{y}}-\tilde{W}_{\bm{x}}^{\dagger}\tilde{W}_{\bm{z}}\right]^{ab}
\,  {\rho^b(\bm{y})} \; .
  \end{equation}
  The  Wilson line $\tilde{W}$ is  given by
  \begin{equation}
  \label{eq:wilsonu}
  \tilde{W}_{\bm{x}} \; = \; P \exp \Big( g \int dx^+ \frac{\delta}{\delta \rho_a(x^+,\bm{x})} \, T^a\Big) \; ,
  \end{equation}
  where now the integration and ordering is in the other light-cone variable $x^+$.
  The above evolution equation thus includes the process of Pomeron splitting by evolving the weight function of the target. The duality can be also rexpressed in term of the symmetry between the projectile and target,
  what is the splitting from the target side, can be interpreted as a Pomeron merging from the projectile side.
  The general Hamiltonian at high energy  should  therefore include both of these processes at the same time. 
  It can be shown \cite{HIMST} that it is entirely expressed in terms of Wilson lines along $x^-$ and $x^+$ directions
 \begin{align}\label{selfdual1}
     H_{\rm eff}
    = \frac{1}{2\pi g^2 N_c}\,
    \int_{\bm{x}}
    {\rm Tr}
    &\big[
   \tilde{V}_\infty(\partial^i \tilde{W}_{\infty})(\partial^i
   \tilde{V}_{-\infty}^{\dagger})\tilde{W}^{\dagger}_{-\infty}+\tilde{V}_\infty
   \tilde{W}_{\infty}(\partial^i \tilde{V}_{-\infty}^{\dagger})
   (\partial^i\tilde{W}^{\dagger}_{-\infty}) \nonumber \\
   &+(\partial^i\tilde{W}^{\dagger}_{-\infty})(\partial^i \tilde{V}_\infty)
   \tilde{W}_{\infty} \tilde{V}_{-\infty}^{\dagger}+\tilde{W}^{\dagger}_{-\infty}(\partial^i \tilde{V}_\infty)
  (\partial^i\tilde{W}_{\infty}) \tilde{V}_{-\infty}^{\dagger}
    \big] \; ,
\end{align}
where the infinity limits in the Wilson lines $\tilde{V}$ and $\tilde{W}$ correspond to the limits in the $x^+$ and $x^-$ lightcone
variables correspondingly. This Hamiltonian is reminiscent of the nonlinear sigma model for the effective theory at high energy,  proposed  in \cite{VV}.  The important point to note here however, is that $\tilde{V}$ and $\tilde{W}$ are  interpreted as operators which are the functions of fields and derivatives of fields, respectively, with very nontrivial commutation relations. The effective Hamiltonian (\ref{selfdual1})
has an interesting symmetry property, namely it is invariant under the following transformation
 \begin{align} \tilde{W}_{-\infty} \to \tilde{V}_{-\infty},  \qquad
   \tilde{V}_{-\infty} \to \tilde{W}_{\infty}^{\dagger},  \qquad
   \tilde{W}_{\infty}^{\dagger} \to \tilde{V}_{\infty}^{\dagger},  \qquad
   \tilde{V}_{\infty}^{\dagger} \to \tilde{W}_{-\infty},
    \end{align}
   which is refferred to as the selfduality property. The effective Hamiltonian reduces to JIMWLK (\ref{eq:hjimwlk})
   and its dual  (\ref{eq:hbrem}) when the Wilson lines $\tilde{W}$ and $\tilde{V}$ are expanded to the first nontrivial order in derivatives 
   and fields, respectively. 
   

\begin{theacknowledgments}
The results presented in this talk have been obtained in the collaboration with Y.Hatta, E.Iancu, L.McLerran and
D.Triantafyllopoulos \cite{HIMST}.
 This
research has been  supported by the U. S. Department of Energy,
Contract No. DE-AC02-98CH10886  and by the 
 Polish Committee for
Scientific Research, KBN Grant No. 1 P03B 028 28. 
\end{theacknowledgments}


\end{document}